\documentclass[12pt]{article}
\usepackage{pdproc,epsfig} 

\def\bea{\begin{eqnarray}}
\def\eea{\end{eqnarray}}
\def\beq{\begin{equation}}
\def\eeq{\end{equation}}
\def\bq{\begin{quote}}
\def\eq{\end{quote}}
\def\gappeq{\mathrel{\rlap {\raise.5ex\hbox{$>$}} {\lower.5ex\hbox{$\sim$}}}}
\def\lappeq{\mathrel{\rlap{\raise.5ex\hbox{$<$}} {\lower.5ex\hbox{$\sim$}}}}

\def\be{\begin{equation}}
\def\ee{\end{equation}}
\def\bc{\begin{center}}
\def\ec{\end{center}}
\def\bea{\begin{eqnarray}}
\def\eea{\end{eqnarray}}

\def\gappeq{\mathrel{\rlap {\raise.5ex\hbox{$>$}} {\lower.5ex\hbox{$\sim$}}}}
\def\lappeq{\mathrel{\rlap{\raise.5ex\hbox{$<$}} {\lower.5ex\hbox{$\sim$}}}}
\newcommand{\bac}{\beq\begin{array}}
\newcommand{\eac}{\end{array}\eeq}
\newcommand{\ba}{\begin{array}}
\newcommand{\ea}{\end{array}}
\newcommand{\beaa}{\begin{eqnarray*}}
\newcommand{\eeaa}{\end{eqnarray*}}

\begin{document}

\vspace*{-1cm}
\phantom{hep-ph/***}

\hfill{RM3-TH/11-7}
\hfill{CERN-PH-TH/2011-169}

\vskip 2.5cm
\renewcommand{\thefootnote}{\alph{footnote}}
  
\title{PERSPECTIVES IN NEUTRINO PHYSICS}

\author{GUIDO ALTARELLI}

\address{Dipartimento di Fisica, Universita' di Roma Tre\\
Rome, Italy\\
and\\
CERN, Department of Physics, Theory Division \\  
CH-1211 Gen\`eve 23, Switzerland\\
{\rm E-mail: guido.altarelli@cern.ch}}

\abstract{This is a Concluding Talk, not a Summary of the Conference. I will discuss some of the
highlights that particularly impressed me (a subjective choice) and make some comments on the status and the prospects of neutrino mass and mixing.}

\normalsize\baselineskip=15pt

\vskip 2cm

\section{Introduction} 

So far the main theoretical lessons from $\nu$ mass and mixing \cite{njp} (see the talks by A. McDonald, E. Lisi) are that
$\nu$Õs are not all massless but their masses are very small; probably their masses are small because $\nu$Õs are Majorana fermions with masses inversely proportional to the large scale M of interactions that violate lepton number (L) conservation. From the see-saw formula, the observed atmospheric oscillation frequency and a Dirac mass of the order of the Higgs VEV $M \sim m_{\nu R}$ is empirically close to $10^{14}-10^{15}$ GeV $\sim M_{GUT}$, so that $\nu$ masses fit well in the SUSY GUT picture.  Decays of $\nu_R$ with CP and  L violation can produce a sizable B-L asymmetry compatible with baryogenesis via leptogenesis (see the talk by P. Di Bari). There is still no direct proof that neutrinos are Majorana fermions: detecting neutrino-less double beta decay ($0 \nu \beta \beta $) would prove that $\nu$Õs are Majorana particles and that L is violated (talks by F. Iachello, J. Gomez-Cadenas, C. Brofferio and S. Shonert). It also appears that $\nu$Õs are not a significant component of dark matter in Universe (talks by A. Melchiorri, G. Gelmini).

\section{Experimental Highlights}

On the experimental side the main developments were the first results from T2K and the coming back of sterile neutrinos (one also talks, without a solid basis, of CPT violation and of non standard interactions). As well known, the T2K run was suddenly interrupted by the devastating earthquake that hit Japan on March 11, just minutes away from the scheduled presentation of the first T2K data. Andrea Rubbia presented here the experiment, the analysis and the first results and I will come back to this later. Actually while this writeup was in preparation T2K released the first publication on their data  \cite{t2k}, reporting a 2.5$\sigma$ signal  for $\sin^2{2\theta_{13}}$, a very important development that indicates for $\theta_{13}$ a value close to the previous upper bound, of the order of the Cabibbo angle $\theta_C$. I will comment on the implications of the T2K result in this article. 

On the evidence for sterile neutrinos a number of hints have been reported at this Conference. They do not make yet a clear evidence but certainly pose an experimental problem that needs clarification. First, there is the MiniBooNE experiment (presented by G.Mills) that in the antineutrino channel reports an excess of events supporting the LSND oscillation signal (originally observed with antineutrinos). The MiniBooNE best fit point falls in an excluded area but there is an overlap with the LSND signal in an allowed region. In the neutrino channel MiniBooNE did not observe a signal in the LSND domain. However, in these data there is a unexplained excess at low energy over the (reliably?) estimated background.  In the neutrino data sample, for the search of a LSND-like signal, only the events with neutrino energy above a threshold value $E_{th}$ were used, leaving the issue of an explanation of the low energy excess unanswered. In the antineutrino channel most of the support to the LSND signal appears to arise from an excess above $E_{th}$ but quite close to it, so that there is, in my opinion, some room for perplexity. Then there is the reactor anomaly: a reevaluation of the reactor flux \cite{flux}, presented at this Conference by T. Lasserre, produced an apparent gap between the theoretical expectations and the data taken at small distances from the reactor ($\lappeq$ 100 m). The discrepancy is of the same order of the quoted systematic error whose estimate, detailed in the paper, should perhaps be reconsidered. Similarly the Gallium anomaly \cite{gal} depends on the assumed cross-section which could be questioned. The reactor anomaly and the Gallium anomaly do not really agree on the oscillation parameters that they point to: the $\Delta m^2$ values are compatible but the central values of $\sin^2{2\theta}$ differ by about an order of magnitude, with Gallium favouring the larger angle. As reported by A. Melchiorri, cosmological data appear compatible with at most the existence of 1 sterile neutrino (the most stringent bounds arising form nucleosynthesis). Over all, only a small leakage from active to sterile neutrinos is allowed by present neutrino oscillation data, as discussed by C. Giunti. If all the indications listed above were confirmed (it looks unlikely) then 1 sterile neutrino would not be enough and at least 2 would be needed with sub-eV masses (as presented by C.Giunti and T. Schwetz). Establishing the existence of sterile neutrinos would be a great discovery. In fact a sterile neutrino is an exotic particle not predicted by the most popular models of new physics. A sterile neutrino is not a 4th generation neutrino: the latter is coupled to the weak interactions (it is active) and heavier than half the Z mass. A sterile neutrino would probably be a remnant of some hidden sector. The issue is very important so that new and better experimental data are badly needed. MiniBooNE will present new results in the summer (as stated by G. Mills). As presented by Carlo Rubbia, Icarus (the apparatus and the technique were discussed by F. Pietropaolo) proposes a new experiment at CERN with two Argon detectors, a close one, at 150 m,  of 150 tons (to be built) and a far one, at 800 m, of 600 t  (to be carried from Gran Sasso where it is now located). This would be the dream experiment for sterile neutrinos but the cost and time scale are relatively large.

Besides MiniBooNE a not yet significant hint of a difference between neutrinos and antineutrinos oscillation parameters
is also reported by MINOS and presented by L. Corwin. This could in principle be interpreted as CPT violation, but also as a matter effect, especially in the presence of non standard interactions (see the talk by M. Maltoni). Normally we expect that this hint of a discrepancy will go away when MINOS will soon present the results with more statistics from the continuation of the run. Actually, as reported by J. Wilkes, a difference between the oscillation parameters of neutrinos and antineutrinos is not supported by SuperKamiokande (which also poses strong contraints on sterile neutrinos). Neutrinos and anti-neutrinos are distinguished on a statistical basis from the angular distributions of the detected charged leptons. 

In neutrino oscillations the leakage from the three active species towards the sterile neutrinos is any case small and, in fact, the best established oscillation phenomena are well described in terms of 3-neutrino models. Recently the main developments have been the T2K and MINOS results on $\theta_{13}$.   The T2K result \cite{t2k}, based on the observation of 6 electron events when $1.5\pm0.3$ are expected for $\theta_{13}=0$, is converted into a confidence interval $0.03(0.04)\le \sin^2{2\theta_{13}} \le 0.28(0.34)$ at $90\%$ c.l. for $\sin^2{2\theta_{23}}=1$, $|\Delta m^2| = 2.4 10^{-3} eV^2$, $\delta_{CP} = 0$ and for normal (inverted) neutrino mass hierarchy. Recently the MINOS Collaboration released \cite{min} their corresponding $90\%$ c.l. range
as $0(0)\le \sin^2{2\theta_{13}} \le 0.12(0.19)$. At the moment of writing this article (July 2011), the best values of the oscillation parameters, including the recent T2K and MINOS results, obtained from a 3 active neutrino analysis in Ref. \cite{fogl}, are reported in Table 1. Taken at face value the combined value of $\sin^2{\theta_{13}}$ is about $3\sigma$ away from zero. This result has very important implications on neutrino oscillation physics. First, it is very good news for the possibility of detecting CP violation in neutrino oscillations. Second, the relatively large central value for $\sin^2{\theta_{13}}$ in Table 1 has a strong impact in discriminating models of neutrino mixing. In fact, this corresponds to $\sin{\theta_{13}}\sim 0.158$, which is comparable to $\lambda_C= \sin{\theta_C} \sim 0.226$.

\begin{table}[h]
\begin{center}
\begin{tabular}{|c|c|}
\hline
&\\[-4mm]
  & ref. \cite{fogl}   \\[2mm]
\hline
&\\[-4mm]
$\delta m^2/10^{-5} eV^2$ &$7.58^{+0.22}_{-0.26}~$ \\[2mm]
\hline
&\\[-4mm]
$\Delta m^2/10^{-3} eV^2$ &$2.35^{+0.12}_{-0.21}~~$ \\[2mm]
\hline
&\\[-4mm]
$\sin^2\theta_{12}$ &$0.312^{+0.017}_{-0.016}~$ \\[2mm]
\hline
&\\[-4mm]
$\sin^2\theta_{23}$ &$0.42^{+0.08}_{-0.03}~$ \\[2mm]
\hline
&\\[-4mm]
$\sin^2\theta_{13}$ &$0.025\pm0.007$  \\[2mm]
\hline
  \end{tabular}
\end{center}
\begin{center}
\begin{minipage}[t]{12cm}
\caption{\label{table:OscillationData} Results of a recent fit to the neutrino oscillation parameters (1$\sigma   $ errors are shown).}
\end{minipage}
\end{center}
\end{table}

For the near future the most important experimental challenges on neutrino oscillation experiments are more precise measurements of the absolute scale of neutrino mass, the accurate determination of $\theta_{13}$ (for the reactor experiments Double CHOOZ, Daya Bay and RENO see the talks by A. Cabrera Serra, Y. Wang and S-B Kim, respectively) and of the shift from maximal of $\theta_{23}$, the fixing of the sign of $\Delta m^2_{23}$ (normal or inverse hierarchy) (see the talk by A. Sousa on No$\nu$A), the detection of CP violation in $\nu$ oscillations (for the future of $\nu$ oscillations see the talks by I. Efthymiopoulos, S. Parke, E. Wildner, K. Long, A. Rubbia, M. Bishai, J. Alonso etc). Related to neutrino physics is the issue of the non conservation of the separate e, $\mu$ and $\tau$ lepton numbers. We are all waiting with great interest for the results of the MEG experiment at PSI (presented by A. Baldini) which is taking data with the goal of bringing the sensitivity to the branching ratio of $\mu \rightarrow e \gamma$ from the present level of $10^{-11}$ down to $10^{-12}-10^{-13}$. This is a channel where a positive signal at this level of sensitivity is expected in many plausible extensions of the Standard Model (see the talk by P. Paradisi).

\section{Models of Neutrino Mixing}

\begin{figure}[]
\centering
$$\hspace{-4mm}
\includegraphics[width=10.0 cm]{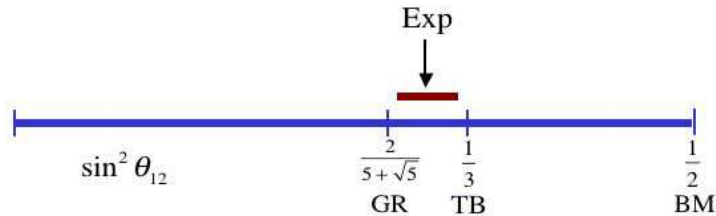}$$
\caption[]{The values of $\sin^2{\theta_{12}}$ for TB or GR or BM mixing are compared with the data }
\end{figure}

To illustrate the impact of the new results on $\theta_{13}$ on models of neutrino mixing (see the talk by S. F. King) we consider the case of models based on discrete flavour groups that have received a lot of attention in recent years \cite{rmp}. There are a number of special mixing patterns that have been studied in this context. These mixing matrices all have $\sin^2{\theta_{23}}=1/2$, $\sin^2{\theta_{13}}=0$  and differ by the value of $\sin^2{\theta_{12}}$  (see Fig. 1). The observed value of $\sin^2{\theta_{12}}$ \cite{fogl}, the best measured mixing angle,  is very close, from below, to the so called Tri-Bimaximal (TB) value \cite{hps} which is $\sin^2{\theta_{12}}=1/3$. Alternatively it is also very close, from above, to the Golden Ratio (GR) value \cite{fp} which is $\sin^2{\theta_{12}}=\frac{1}{\sqrt{5}\phi} = \frac{2}{5+\sqrt{5}}\sim 0.276$, where $\phi= (1+\sqrt{5})/2$ is the GR. Thus, a possibility is that one of these coincidences is taken seriously and this leads to models where TB or GR mixing is naturally predicted as a good first approximation. Here I will mainly refer to TB mixing which is the most studied first approximation to the data. One considers models where TB mixing is implied by the underlying dynamics and provides a leading order approximation corrected by non leading effects. Alternatively one can assume that this agreement of the data with TB mixing is accidental. Indeed there are many models that fit the data and yet TB mixing does not play a role in their architecture. For example, in ref.(\cite{alro}) there is a list of Grand Unified SO(10) models with excellent fits to the  neutrino mixing angles although most of them have no relation with TB mixing. If instead we assume that TB mixing has a real dynamical meaning then it is important to consider models that naturally lead to TB mixing. In a series of papers (for a detailed list of references see \cite{rmp}) it has been pointed out that a broken flavour symmetry based on the discrete
group $A_4$ appears to be particularly suitable to reproduce this specific mixing pattern in leading order (in the case of GR mixing the simplest choice is the group $A_5$ \cite{fp}). We recall that $A_n$ is the group of even permutations of n objects (n!/2 elements). Other
solutions for TB mixing based on alternative discrete or  continuous flavour groups have also been considered, but the $A_4$ models have a very economical and attractive structure, e.g. in terms of group representations and of field content. 
In most of the models $A_4$ is accompanied by additional flavour symmetries, either discrete like $Z_N$ or continuous like U(1), which are necessary to eliminate unwanted couplings, to ensure the needed vacuum alignment and to reproduce the observed mass hierarchies. Given the set of flavour symmetries and having specified the field content, the non leading corrections to the TB mixing arising from higher dimensional effective operators can be evaluated in a well defined expansion. In the absence of specific dynamical tricks, in a generic model, all the three mixing angles receive corrections of the same order of magnitude. Since the experimentally allowed departures of $\theta_{12}$ from the TB value $\sin^2{\theta_{12}}=1/3$ are small, numerically at most of $\mathcal{O}(\lambda_C^2)$, it follows that both $\theta_{13}$ and the deviation of $\theta_{23}$ from the maximal value are expected in these models to also be at most of $\mathcal{O}(\lambda_C^2)$ (note that $\lambda_C$ is a convenient hierarchy parameter not only for quarks but also in the charged lepton sector with $m_\mu/m_\tau \sim0.06 \sim \lambda_C^2$ and $m_e/m_\mu \sim 0.005\sim\lambda_C^{3-4}$). A value of $\theta_{13} \sim \mathcal{O}(\lambda_C^2) \sim \mathcal{O}(0.05)$ is now rather marginal in view of the T2K result. Thus models based on TB or GR mixing are now somewhat disfavoured. It is true that one can introduce some additional theoretical input to suppress the value of $\theta_{13}$. In the case of $A_4$, an example is provided by the model of ref. \cite{linx}, formulated before the T2K result was known, and also the modified $A_4$ model of ref. \cite{A4mod}.

The new results showing that probably $\theta_{13}$ is near its former upper bound could be interpreted as an indication that the agreement with the TB or GR mixing is accidental. Then a scheme where instead the Bimaximal (BM) mixing is the correct first approximation modified by terms of $\mathcal{O}(\lambda_C)$ could be relevant. In BM mixing $\theta_{12}$ and $\theta_{23}$ are both maximal while $\theta_{13}=0$ (see. Fig. 1). This is in line with the well known empirical observation that $\theta_{12}+\theta_C\sim \pi/4$, a relation known as quark-lepton complementarity \cite{compl}, or similarly $\theta_{12}+\sqrt{m_\mu/m_\tau} \sim \pi/4$. No compelling model leading, without parameter fixing, to the exact complementarity relation has been produced so far. Probably the exact complementarity relation becomes more plausible if replaced with $\theta_{12}+\mathcal{O}(\theta_C)\sim \pi/4$ or $\theta_{12}+\mathcal{O}(m_\mu/m_\tau)\sim \pi/4$ (which we could call "weak" complementarity). One can think of models where, because of a suitable symmetry,  BM mixing holds in the neutrino sector at leading order and the necessary, rather large, corrective terms for $\theta_{12}$ arise from the diagonalization of charged lepton masses \cite{compl}. These terms of order $\mathcal{O}(\lambda_C)$ from the charged lepton sector would then generically also affect $\theta_{13}$ an the resulting value could well be compatible with the T2K result.
Along this line of thought, we have used the expertise acquired with non Abelian finite flavour groups to construct a model \cite{S4us} based on the permutation group $S_4$ which naturally leads to the BM mixing at leading order. We have adopted a supersymmetric formulation of the model in 4 space-time dimensions. The complete flavour group is $S_4\times Z_4 \times U(1)_{FN}$. In leading order, the charged leptons are diagonal and hierarchical and the light neutrino mass matrix, after see-saw, leads to the exact BM mixing. The model is built in such a way that the dominant corrections to the BM mixing, from higher dimensional operators in the superpotential,  only arise from the charged lepton sector at next-to-the-leading-order and naturally inherit $\lambda_C$ (which fixes the charged lepton mass hierarchies) as the relevant expansion parameter. As a result the mixing angles deviate from the BM values by terms of  $\mathcal{O}(\lambda_C)$ (at most), and weak complementarity holds. A crucial feature of the model is that only $\theta_{12}$ and $\theta_{13}$ are corrected by terms of $\mathcal{O}(\lambda_C)$ while $\theta_{23}$ is unchanged at this order (which is essential for a better agreement of the model with the present data). Recently the model was extended to include quarks in a $SU(5)$ Grand Unified version \cite{melo}.

We now briefly turn to models that do not take seriously any of the coincidences described above (the proximity of the data to the TB or GR patterns or the quark-lepton complementarity: they cannot all be true and it is possible that none of them is true) and are therefore based on a less restrictive flavour symmetry. It is clear that the T2K hint that $\theta_{13}$ may be large is great news for the most extreme position of this type, which is "anarchy" \cite{hmw}: no symmetry at all in the lepton sector, only chance. This view predicts generic mixing angles, so the largest $\theta_{23}$ should be different than maximal and the smallest $\theta_{13}$ should be as large as possible within the experimental bounds. Anarchy can be formulated in a $SU(5) \bigotimes U(1)$ context by taking different Froggatt-Nielsen charges only for the $SU(5)$ tenplets (for example 10: (3,2,0) where 3 is the charge of the first generation, 2 of the second, zero of the third) while no charge differences appear in the $\bar 5$: $\bar 5$: (0,0,0). This assignment is in agreement with the empirical fact that the mass hierarchies are more pronounced for up quarks in comparison with down quarks and charged leptons. In a non see-saw model, with neutrino masses dominated by the contribution of the dimension-5 Weinberg operator, the $\bar 5$ vanishing charges directly lead to random neutrino mass and mixing matrices. In anarchical see-saw models also the charges of the SU(5) singlet right-handed neutrinos must be undifferentiated. Anarchy can be mitigated by assuming that it  only holds in the 2-3 sector: e.g $\bar 5$: (2,0,0) with the advantage that the first generation masses and the angle $\theta_{13}$ are naturally small. In models with see-saw one can alternatively play with the charges for the right-handed SU(5) singlet neutrinos. If, for example, we take 1: (1, -1, 0), together with $\bar 5$: (2,0,0),  it is possible to get a normal hierarchy model with $\theta_{13}$ small and also with $r = \Delta m^2_{solar}/\Delta m^2_{atm}$ naturally small (see, for example, ref. \cite{afm}). In summary anarchy and its variants, all based on chance, offer a rather economical class of models that are among those encouraged by the new $\theta_{13}$ result.

\section{Conclusion}

I imagine that by the next edition of this by now classic Conference, in 2013, we will know the value of $\theta_{13}$ with a good accuracy, from the continuation of T2K and from the start of the reactor experiments Double CHOOZ, Daya Bay and RENO. Many existing models will be eliminated and the surviving ones will be updated to become more quantitative in order to cope with a precisely known mixing matrix. A sizable $\theta_{13}$ will encourage the planning of long baseline experiments for the detection of CP violation in neutrino oscillations. Along the way the important issue of the existence of sterile neutrinos must be clarified. The on going or in preparation experiments on the absolute value of neutrino masses, on $0\nu \beta \beta$, on $\mu \rightarrow e \gamma$, on the search for dark matter etc can also lead to extremely important developments in the near future. So this field is very promising and there all reasons to expect an exciting time ahead of us.

As a last speaker, on behalf of all partecipants I would like to most warmly thank the Organisers of this Conference. But I am sorry that this is the first NeuTel without the charming presence of Milla Baldo Ceolin.
So I extend our best greetings and warmest wishes to Milla.

\section{Acknowledgments}
It is a very pleasant duty for me to most warmly thank Mauro Mezzetto and the Organising Committee for their kind invitation and for the great hospitality offered to all of us in Venice. I am also glad to acknowledge interesting discussions on this subject with Ferruccio Feruglio, Davide Meloni and Luca Merlo.

\end{document}